# Change Point Analysis of Multivariate Data: Using Multivariate Rank-based Distribution-free Nonparametric Testing via Measure Transportation with Applications in Tumor Microarrays and Dementia

Amanda Ng

The Bronx High School of Science

# Abstract

In this paper, I propose a general algorithm for multiple change point analysis via multivariate distribution-free nonparametric testing based on the concept of ranks that are defined by measure transportation. Multivariate ranks and the usual one-dimensional ranks both share an important property: they are both distribution-free. This finding allows for the creation of nonparametric tests that are distribution-free under the null hypothesis. This method has applications in a variety of fields, including medical diagnostics, and in this paper I implement this algorithm to a microarray dataset for individuals with bladder tumors. Since all individuals in the dataset share similar circumstances, it is anticipated that the change points yielded in each dataset will have approximately the same positions. The algorithm has also been applied to an ECoG snapshot for a patient with epilepsy, with the change points representing temporal transitions between different states. This algorithm has also been described in the context of trajectories of CASI scores by education level and dementia status and individuals with and without dementia. Each change point denotes a shift in the rate of change of Cognitive Abilities score over years, indicating the existence of preclinical dementia.

The algorithm presented here can also be applied to medical condition monitoring, as continuous monitoring requires trend detection in physiological variables, including heart rate, electrocardiogram (ECG), and electroencephalogram (EEG). Trend detection in these variables is necessary in order to study specific medical issues, including understanding brain activity, epilepsy, sleep problems, and MRI interpretation.

Here I will consider rank energy statistics in the context of the multiple change point problem. I will estimate the number of change points and each of their locations within a multivariate series of time-ordered observations. This paper will examine the multiple change point question in a broad setting in which the observed distributions and number of change points are unspecified, rather than assume the time series observations follow a parametric model or there is one change point, as many works in this area assume. The objective here is to create an algorithm for change point detection while making as few assumptions about the dataset as possible.

This algorithm described here is based upon energy statistics and has the ability to detect any distributional change. Presented are the theoretical properties of this new algorithm and the conditions under which the approximate number of change points and their locations can be estimated. This newly proposed algorithm can be used to analyze various datasets, including financial, ECoG, CASI, and microarray data. This algorithm has also been successfully implemented in the R package `recp`, which is available on GitHub. A section of this paper is dedicated to the execution of this procedure, as well as the use of the `recp` package.

**Introduction**

Change point detection is the process of detecting changes in the distribution of time-ordered observations. This arises in medical condition monitoring, speech analysis, climate change detection, and human activity analysis. It is also applied in financial modeling, where assets are traded and models based on historical data are represented as a multivariate time series [20]. Change point analysis can also be used to detect credit card fraud [21] and other anomalies. It can also be used in signal processing, where change point analysis is used to detect notable changes within a stream of images [4].

The algorithm proposed in this paper can also be applied to microarray data. In this paper, I will analyze the results obtained by applying the recp method to data for 57 individuals with bladder tumors. Since all individuals in the dataset share similar circumstances, it is anticipated that the change points yielded in each dataset will have approximately the same positions.

Moreover, this algorithm can also be used in the context of medical condition monitoring. Monitoring patient health involves change point detection and trend detection in a variety of physiological variables, including heart rate, electrocardiogram (ECG), and electroencephalogram (EEG). Research on change point detection also investigates change point detection for a variety of medical issues, including understanding brain activity, epilepsy, sleep problems, and MRI interpretation. Here, the change point algorithm is implemented in an ECoG snapshot for an individual with epilepsy, and each change point detected denotes the brain's temporal transition between states. This algorithm has also been described in the context of trajectories of CASI scores by education level and dementia status and individuals with and without dementia, in which each change point denotes a shift in the rate of change of Cognitive Abilities score over years. The results from this implementation also indicate the existence of preclinical dementia.

Although change point analysis is significant in a variety of fields, the methodologies currently established often assume a specific number of change points. This assumption is unrealistic, and applications increasingly require detecting changes in multivariate data in which these traditional methods have restricted applicability. To address these limitations, I propose a new methodology in which rank-based energy statistics are used in the context of change point detection with a basis on U-statistics. This is able to consistently estimate the location of an arbitrary number of multiple change point locations.

Change point analysis can be performed in both parametric and nonparametric settings. In parametric analysis, the underlying distributions are assumed to belong to some known family and the likelihood function plays a significant role. Nonparametric approaches, however, are applicable in a wider variety of datasets and fields than their parametric counterparts. They depend significantly on the approximation of density functions and have also been implemented using rank-based distribution-free nonparametric testing [1]. Proposed here is a nonparametric algorithm based on rank-based distribution-free nonparametric testing, which does not involve the complexities correlated with multivariate density approximation.

As previously mentioned, my proposal is exactly distribution-free in finite samples. This is a useful property as it avoids the need to estimate any excess parameters. In addition, the distribution-free property can result in a reduction in the computational complexity in statistical problems. Computing the suggested test statistic is computationally achievable under any dimension and sample size and does not involve adjusting any parameters.

The only condition required of the underlying distributions is that they must be absolutely continuous and no moment conditions are necessary in order to maintain the consistency of the test. This allows for the usage of nonparametric inference under data-generating distributions with heavy tails.

Change point algorithms typically estimate all change points simultaneously or hierarchically. Simultaneous methods often optimize a single objective function. For example, [2] estimates the change point locations by maximizing a likelihood function and [3] performs the same task by minimizing a loss function.

I propose a new method that can detect any distributional change in an independent sequence using ranked energy statistics, and which does not make any assumptions beyond the absolute moment for which it ∈ (0,2). This method is performed in a way that identifies the number and locations of change points concurrently. In the next section I discuss methodology and its properties then present the results of my procedure when applied to real data and concluding comments are in the last section.

**Methodology**

Here I will consider the rank energy statistic, a distribution-free goodness-of-fit measure using E-statistics, in the context of estimating the number and locations of change points. The rank energy statistic itself is used for testing the equality of two multivariate distributions.

I describe the method below, where $\mu_m^X$ and $\mu_n^Y$ represent the empirical distributions on $D_m^X := \{X_1,..., X_m\}$ and $D_n^Y := \{Y_1, ..., Y_n\}$ respectively. Let $\mu_{m,n}^{X,Y} := (m+n)^{-1}(m\mu_n^X + n\mu_n^Y)$ and let $H_{m+n}^d := \{h_1^d, ..., h_{m+n}^d\} \subset [0,1]^d$ designate the fixed sample multivariate ranks.

The empirical distribution on $H_{m+n}^d$ weakly converges to $U^d$ as $\min(m,n) \to \infty$, in which $H_{m+n}^d$ is the d-dimensional Halton sequence for d≥2 and $\{i/(m+n) : 1 \le i \le m+n\}$ for d=1.

$\widehat{R}_{m,n}^{X,Y}(\cdot)$ denotes the joint empirical rank map corresponding to the transportation of $\mu_{m,n}^{X,Y}$ to the empirical distribution $H_{m+n}^d$. The change point detection using a rank energy statistic is therefore defined as:

$$\widehat{\mathcal{E}}(X_n, Y_m; \alpha) = \frac{2}{mn} \sum_{i=1}^{n}\sum_{j=1}^{m} |\widehat{R}_{m,n}^{X,Y}(X_i) - \widehat{R}_{m,n}^{X,Y}(Y_j)|^\alpha - \binom{n}{2}^{-1} \sum_{1 \le i \le k \le n} |\widehat{R}_{m,n}^{X,Y}(X_i) - \widehat{R}_{m,n}^{X,Y}(X_k)|^\alpha$$
$$- \binom{m}{2}^{-1} \sum_{1 \le j \le k \le m} |\widehat{R}_{m,n}^{X,Y}(Y_j) - \widehat{R}_{m,n}^{X,Y}(Y_k)|^\alpha \qquad (1)$$

$$\widehat{Q}(X_n, Y_m; \alpha) = \frac{mn}{m+n} \widehat{\mathcal{E}}(X_n, Y_m; \alpha) \qquad (2)$$

$$(\hat{\tau}, \hat{\kappa}) = \underset{(\tau,\kappa)}{argmax}\ \widehat{Q}(X_\tau, Y_\tau(\kappa); \alpha) \qquad (3)$$

Observe that the right hand side of (1) can be viewed as the rank energy statistic, which is based on Euclidean distances between sample elements. This rank energy statistic can also be viewed as a rank-transformed version of the empirical energy measure as described in [4]. (1) is also described in more detail and more extensively evaluated in [5]. (2) represents the scaled sample empirical divergence measure that leads to a consistent approach for estimating change point locations. Let $Z_1, ..., Z_T \in \mathbb{R}^d$ be an independent sequence of observations and $1 \leq \tau < \kappa \leq T$ be constants. In addition, let the following sets $X_\tau = \{Z_1, Z_2, ..., Z_\tau\}$ and $Y_\tau(\kappa) = \{Z_{\tau+1}, Z_{\tau+2}, ..., Z_\kappa\}$. The change point location $\hat{\tau}$ is then estimated in (3). This procedure can also be seen as a ranked version of the change point detection method described in [6].

All the methods described above have been implemented in the R software. The relevant codes are available in GitHub under the repository found at "**Amanda-Ng/recp**".

**Consistency**

I now present the consistency of the approximated change point locations that are detected by the suggested algorithm. It is presumed that the number of change points detected is constant, though unknown, and the observations have a constant, unknown dimension. Here, I will demonstrate the strong consistency of this estimator in the context of both single and multiple change points in a rescaled time setting.

In the case of a change point at a given location, the two-sample test is statistically significant against any alternatives. $\hat{\tau}$ is a consistent estimator for any single change point location within the setting established. Consistency requires that each cluster's size increase, and each cluster's size does not need to increase at the same rate. However, to evaluate the rate of convergence, supplementary information about the distribution of estimated change points, which depends on the unknown data distribution, is needed.

The consistency established in [18] is not applicable in the context with multiple change points because it assumes that the time series is composed of a piecewise linear function, and thus being consistent only for change point analysis resulting from deviations from expectation.

**An Agglomerative Approach**

The procedure proposed up to this point has considered only the use of a divisive approach, but it is also possible to execute an agglomerative algorithm.

Let the sequence of the observations $Z_1, Z_2, ..., Z_T$ be independent and each observation has a finite $\alpha^{th}$ moment for which $\alpha \in (0,2)$. As opposed to most common agglomerative procedures, the following proposal conserves the time ordering of distributions and the number of change points are approximated through the maximization of the goodness-of-fit statistic.

Assume there is a clustering $C = \{C_1, C_2, ..., C_n\}$ of n clusters, each of which need not consist of a single observation. Let $C_i = \{Z_k, Z_{k+1}, ..._{k+t}\}$ and let $C_j = \{Z_m, Z_{m+1}, ..., Z_{m+s}\}$. In order to preserve the time ordering of observations, let $C_i$ and $C_j$ merge if either k+t+1 = m or m+s+1 = k. In other words, if $C_i$ and $C_j$ are adjacent clusters.

In order to identify which pair of adjacent clusters to merge, a goodness-of-fit statistic must be used, as described below. The statistic is optimized by merging a pair of adjacent clusters that sums in either the largest or smallest decrease in the statistic's value. This approach is repeated and the statistic is recorded each time until all observations form a single cluster. The approximate number of change points is then estimated by a clustering that maximizes the goodness-of-fit statistic over the whole merging sequence.

The goodness-of-fit statistic used here is the distance between adjacent clusters. Assume that $C = \{C_1, C_2, ..., C_n\}$, then $\widehat{S}_n(C; \alpha) = \sum_{i=1}^{n-1} \widehat{Q}(C_i, C_{i+1}; \alpha)$, where $C_i$ and $C_{i+1}$ are adjacent and $\widehat{Q}$ is analogous to equation (2). Initialization of the merging sequence $\{\widehat{S}_k: k = n, ..., 2\}$ is done by calculating $\widehat{Q}$ for all pairs of clusters as done in any agglomerative algorithm.

## `recp`: An R Package for Change Point Analysis of Multivariate Data via Multivariate Rank-based Distribution-free Nonparametric Testing

I introduce the `recp` R package for multiple change point analysis of a multivariate time series via ranked energy statistics. The `recp` package provides a procedure for change point detection that is able to detect any change in distribution within a given time series. Both the number and locations of estimated change points are approximated simultaneously. There are only two assumptions placed on these distributions and they are that the $\alpha^{th}$ moment exists for which $\alpha \in (0, 2]$ and that each observation is independent over time. Changes in distribution are identified by using the ranked energy statistic in equation (1).

There are various available R packages that can be used for change point detection, with each one making its own assumptions about the given time series observed. For example, the `changepoint` R package offers multiple procedures for performing change point detection of a univariate time series. For certain methods, the anticipated computational complexity can be shown to be proportional with the length of the time series. At present, the `changepoint` package is suitable only for detecting changes in the mean or the variance via penalization. The limitation to this algorithm is that it necessitates a user-specified penalty term.

Moreover, the `cpm` R package provides multiple methods for change point detection in a univariate time series, including nonparametric methods, in which general distribution changes are detected, and detecting change points in independent Gaussian data. While this method provides procedures for change detection in univariate time series with arbitrary distribution, this procedure cannot be easily used to detect change points in the joint distribution of multivariate data.

On the other hand, the `bcp` package is used to perform Bayesian single change point analyses of univariate time series. Newer versions of this R package have reduced the computational cost with respect to the length of the time series, although every version of `bcp` is designed to detect change points in only the mean of independent Gaussian distributions.

Another example is the `strucchange` package. While it offers tools for detecting change within linear regression models, many of the tools detect no more than one change within a regression model. There are other methods in this package, however, that can be used in settings in which multiple change points exist. For a given number of change points, this algorithm returns the change point estimations that minimize the residual sum of squares.

Above I have introduced the rank energy statistic– the fundamental divergence measure used in change point analysis. Below is more information about the `recp` package and each section's function. Following are examples of the package's procedures applied to real datasets. Lastly, the `recp` package can be obtained in GitHub.

The `recp` package is designed to address many of the limitations within currently available change point packages. It is able to perform multiple change point analysis for a multivariate time series while detecting all departures from the null hypothesis. This algorithm assumes that each of the observations is independent with a finite $\alpha^{\text{th}}$ absolute moment for which $\alpha \in (0, 2]$.

The required packages for `recp` are first uploaded in the following:

```
require(clue, quietly=T)
require(energy, quietly = T)
require(randtoolbox, quietly = T)
require(pracma, quietly = T)
require(kernlab, quietly = T)
require(crossmatch, quietly = T)
require(HHG, quietly = T)
require(gTests, quietly = T)
require(ramify)
```

Data is then generated with the following code in R:

```
m=200
n=200
data1=cbind(rcauchy(m,0,1),rcauchy(m,0,1))
data2=cbind(rcauchy(n,0.5,1),rcauchy(n,0,1))
```

The rank energy statistic, also known as the distribution-free multivariate two-sample testing represented in (1), is calculated:

```
computestatistic=function(x,y,m=nrow(x),n=nrow(y),dim=ncol(x),gridch=torus(m+n,dim))
{
  comdata=rbind(x,y)
  distmat=matrix(0,nrow=m+n,ncol=m+n)
  for(i in 1:(m+n))
    distmat[i,]=apply((comdata[i,]-t(gridch)),2,Norm)^2
  assignmentFUN=solve_LSAP(distmat)
  assignmentSOL=cbind(seq_along(assignmentFUN),assignmentFUN)
  randenergySTAT=eqdist.etest(gridch[assignmentSOL[,2],],sizes = c(m,n), R=1)
  return(randenergySTAT$statistic)
```

Furthermore, let the following denote the scaled sample measure of divergence that corresponds to (2):

```
EqDisttest=((m+n)/(m*n))(randenergySTAT)
```

The change point locations are finally estimated below, which corresponds to (3):

```
changept=argmax(EqDisttest)}
```

**Applications**

In this section, I will analyze the results obtained by applying the change point algorithm method to 4 datasets. It is first applied to the microarray data from [7]. In this dataset, presented are records of the copy-number variations for various individuals with bladder tumors. Next, this method is used in the context of an ECoG snapshot for an individual with epilepsy from [22] and a CASI score trajectories for individuals with different education levels and dementia status [24]. This procedure is also used in a dataset of financial time series consisting of weekly log returns of the companies that constitute the Dow Jones Industrial Average [25].

**Microarray Data**

This microarray dataset contains the data for 57 individuals with bladder tumors. Since each individual has the same condition, it is expected that the change point locations are nearly the same in each microarray set. The approach proposed by [7] assumes that each microarray set could be modeled by a piecewise constant function and is therefore concentrated in the mean. To contrast, the change point method is able to not only detect changes in the mean, but also other changes including those in variability [6].

All individuals from the dataset in [7] for which more than 7% of the values were missing were removed, and the missing values were replaced by the average of the adjacent values. 43 individuals were left, and this procedure estimated 35 change points

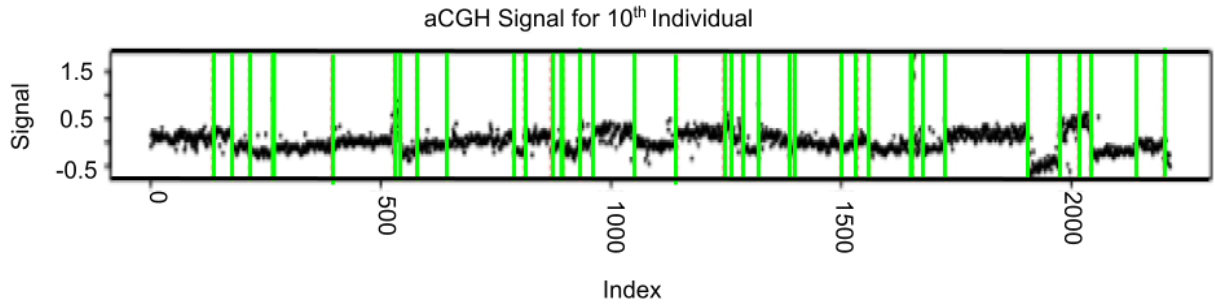

Figure 1: The normalized relative array CGH (aCGH) signal for the tenth individual in the dataset with a bladder tumor. The estimated change point locations are indicated by the green vertical lines.

**Epileptic Activity**

In this section, the change point algorithm is applied to ECoG data to identify sections of brain activity representing the different states of the epileptic brain. The data presented is an ECoG recording from 154 electrodes at 1,000 Hertz from a patient with epilepsy. Figure 2 displays a snapshot of 10 seconds of activity from 4 channels, which each corresponds to an electrode placed between the temporal (T) and parietal (P) lobes of the brain. There are a total of 500 samples from all 4 channels, which are assumed to be independent of each other. The algorithm presented detected approximately 10 change points [22].

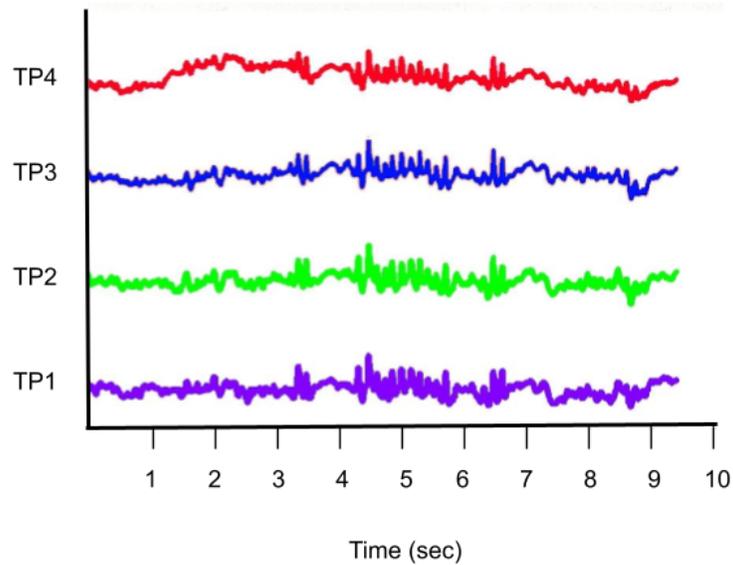

Figure 2: Snapshot of ECoG activity from 4 channels in a 10-second window

**CASI Score and Education Level**

     Here, the change point detection algorithm is used in the context of multiple individuals with different dementia status and education level. The data presented in Figure 3 displays the Cognitive Abilities Screening Instrument (CASI) Score for these individuals over different ages. In individuals who do not have dementia, the CASI scores are 90 and 93 at age 70 for those with lower and higher-level education, respectively. For individuals with dementia, the change points exist at the age of 73 for individuals with lower-level education and 85 for individuals with higher-level education [23]. However, the change point for those with higher-level education precedes a faster rate of cognitive abilities scores decline compared to individuals with dementia who have lower-level education [24].

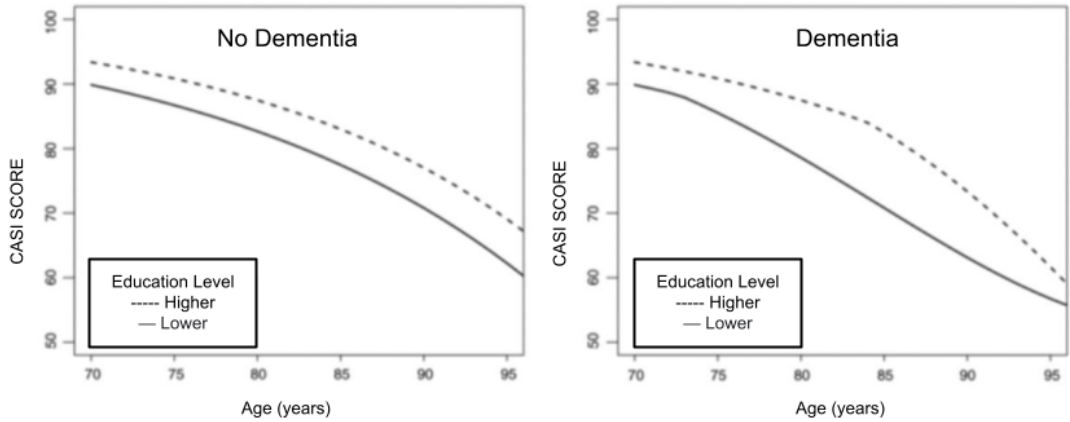

Figure 3: CASI Score trajectories given education level and dementia status in individuals with and without dementia.

**Financial Data**

Here are the weekly log returns considered for the companies that constitute the Dow Jones Industrial Average. The time period analyzed is April 1990 to January 2012 with 1,140 observations. However, the time series for Kraft Foods Inc. does not encompass this entire period, so it will not be considered in this analysis [6].

The proposed method identified change points at these following dates: 07/13/1998, 03/24/2003, 09/15/2008, and 05/11/2009. The change point at 09/15/2008 corresponds to the filing of the Lehman Brothers bankruptcy. The change point at 05/11/2009 corresponds to the release of the results from the Supervisory Capital Asset Management Program [25].

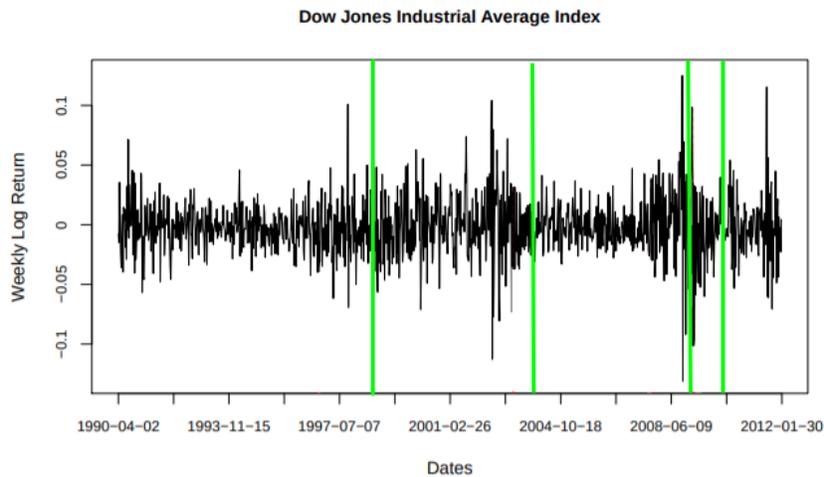

Figure 4: Weekly log returns for the Dow Jones Industrial Average index from April 1990 to January 2012. The green lines indicate the locations of estimated change points at 07/13/1998, 03/24/2003, 09/15/2008, and 05/11/2009.

**Discussion**

I have developed a framework for change point detection via multivariate distribution-free nonparametric testing using the method of multivariate ranks with a basis in the theory of optimal transportation. I have illustrated this general approach to distribution-free multivariate two-sample testing in the context of change point analysis in which I evaluated the number and locations of multiple change points. This proposed test is computationally feasible, finite sample distribution-free, and consistent against alternatives under minimal assumptions. In this process I have also derived results on the asymptotic constancy of optimal transport maps– also known as the multivariate ranks used in (1).

A natural future research direction is to investigate the consistency against local alternatives (e.g. [8]). I believe that my proposed general framework can be used in the context of other time series in which the distribution-free tests are in multivariate nonparametric settings beyond those discussed in this paper. I hope that more of such change point detections via multivariate rank-based distribution-free tests will be studied in the future.

I have presented a method to perform multiple change point analysis of an independent sequence of multivariate distributions via rank energy statistics. This method is able to detect any kind of divergence from $H_0$ and no assumptions are made beyond the existence of the $\alpha$th absolute moment for some $\alpha \in (0,2)$. The proposed method is able to estimate both the number of change points and their respective locations, consequently removing the need for previous knowledge or supplementary analysis, unlike the methods proposed in [1], [2], and [3]. Moreover, this benefit does not come at the expense of computational practicality.

The divisive and agglomerative versions of this procedure have been described. In the divisive version, the statistical significance of each hierarchically estimated change point is hierarchically tested. In the agglomerative version, the goodness-of-fit statistic is optimized. The divisive procedure is preferred in practice as its consistency has been established, although its computational complexity is dependent upon the number of change points approximated.

The `recp` package is able to perform multiple change point analysis of multivariate data. It is able to perform analysis to determine the number and location of change points without user input. The only parameter established by the user is **α**, and if **α** lies within (0,2), then the methods provided in `recp` are able to detect any distributional change within the series provided. Through the provided examples and applications to real data, I observe that the `recp` approach obtains reasonable estimates for the locations of the change points.